\documentclass[letterpaper,titlepage,11pt]{article}

\usepackage[a4paper]{geometry}
\usepackage{amssymb,amsmath,amsthm,amscd}
\usepackage{tensor}
\usepackage{mathtools}
\usepackage{enumerate}
\usepackage[english]{babel}
\usepackage{comment}
\usepackage[numbers,sort&compress]{natbib}
\usepackage{bm}
\usepackage[usenames,dvipsnames]{xcolor}

\usepackage[T1]{fontenc}
\usepackage[utf8]{inputenc}
\usepackage{lmodern}

\usepackage[colorlinks=true,urlcolor=blue,citecolor=magenta]{hyperref}
\usepackage{amssymb,amsmath,amsfonts}
\usepackage{epsfig}
\usepackage{graphicx}
\usepackage{epstopdf}
\usepackage{caption}
\usepackage{subcaption}
\usepackage{amscd}
\usepackage{stmaryrd}
\usepackage{amsthm}
\usepackage{latexsym}
\usepackage{amsbsy}
\usepackage[english]{babel}
\usepackage{psfrag}
\usepackage{tabularx}

\allowdisplaybreaks[1]

\def\clock{{\count0=\time
           \divide\count0 60
           \ifnum\count0<10 0\fi\the\count0
           \multiply\count0 -60 \advance\count0 \time
           :\ifnum\count0<10 0\fi \the\count0
         }}
\newcommand{\timestamp}{{\small\vbox{\hbox{\tt\jobname.tex}
\hbox{\the\day/\the\month/\the\year, \clock}}}}

\newcommand{\CL}{\mathcal{L}}
\newcommand{\CO}{\mathcal{O}}
\newcommand{\CN}{\mathcal{N}}

\newcommand{\C}{\mathbb{C}}
\newcommand{\R}{\mathbb{R}}

\newcommand{\ads}{\mbox{AdS}}

\newcommand{\nn}{\nonumber}
\newcommand{\spa}{\quad , \quad}

\hypersetup{   colorlinks=true,  citecolor=blue, linkcolor=blue, urlcolor=red }

\begin{document}

\numberwithin{equation}{section}

\begin{titlepage}
\rightline{\vbox{   \phantom{ghost} }}

 \vskip 1.8 cm
\begin{center}
{\LARGE \bf
Strings with Non-Relativistic Conformal Symmetry and Limits of the AdS/CFT Correspondence}
\end{center}
\vskip .5cm

\title{}
\date{\today}
\author{Troels Harmark, Jelle Hartong, Lorenzo Menculini, Niels Obers, Ziqi Yan}

\centerline{\large {{\bf Troels Harmark$^1$, Jelle Hartong$^2$, Lorenzo Menculini$^{1,3}$}}}
\vskip 0.2cm
\centerline{\large {{\bf Niels A. Obers$^1$, Ziqi Yan$^4$}}}

\vskip 1.0cm

\begin{center}

\sl $^1$ The Niels Bohr Institute, University of Copenhagen\\
Blegdamsvej 17, DK-2100 Copenhagen \O, Denmark\\
\sl $^2$ School of Mathematics and Maxwell Institute for Mathematical Sciences\\
University of Edinburgh, Peter Guthrie Tait road, Edinburgh EH9 3FD, UK\\
\sl $^3$ Dipartimento di Fisica e Geologia, Universit\`a di Perugia,\\
I.N.F.N. Sezione di Perugia,\\
Via Pascoli, I-06123 Perugia, Italy\\
\sl $^4$ Perimeter Institute for Theoretical Physics\\
31 Caroline St N, Waterloo, ON N2L 6B9, Canada
\vskip 0.4cm

\end{center}

\vskip 1.3cm \centerline{\bf Abstract} \vskip 0.2cm \noindent
We find a Polyakov-type action for strings moving in a torsional Newton-Cartan geometry. This is obtained by starting with the relativistic Polyakov action and fixing the momentum of the string along a non-compact null isometry. For a flat target space, we show that the world-sheet theory becomes the Gomis--Ooguri action. From a target space perspective these strings are non-relativistic but their world-sheet theories are still relativistic. We show that one can take a scaling limit in which also the world-sheet theory becomes non-relativistic with an infinite-dimensional symmetry algebra given by the Galilean conformal algebra. This scaling limit can be taken in the context of the AdS/CFT correspondence and we show that it is realized by the `Spin Matrix Theory' limits of strings on $\ads_5\times S^5$. 
Spin Matrix theory arises as non-relativistic limits of the AdS/CFT correspondence close to BPS bounds. The duality between non-relativistic strings and Spin Matrix theory provides a holographic duality of its own and points towards a framework for more tractable holographic dualities whereby non-relativistic strings are dual to near BPS limits of the dual field theory.


%

\end{titlepage}

\newpage
\tableofcontents


\section{Introduction}

One of the most appealing and successful approaches towards a consistent theory of quantum gravity is string theory. The celebrated discovery
of the AdS/CFT correspondence relating string theory on anti-de Sitter backgrounds to conformal field theories in one dimension less, has provided
a further promising arena to address questions related to quantum gravity, such as the microscopic description of black holes and the information paradox. In these developments the quest to understand quantum gravity has been guided by approaching it from relativistic classical gravity and/or quantum field theory. However, an alternative route is to first consider the quantization of non-relativistic gravity as a step towards (relativistic) quantum gravity. This third path has been much less appreciated, in part since already the classical description of such a theory has not been fully understood%
\footnote{See however the recent work \cite{Hansen:2018ofj} which presents an action principle for Newtonian gravity,  but also goes beyond by  including the effects of strong gravitational fields in non-relativistic gravity.}, nor how such a theory connects to string theory. 
The study of non-relativistic string theory in this paper is motivated by pursuing this latter route, and in particular by applying it to the realm of the AdS/CFT correspondence. 

A natural setting in which to expect a connection between non-relativistic gravity, string theory and holography is to consider the Spin Matrix Theory (SMT) limits of  \cite{Harmark:2014mpa}. These are tractable limits of the AdS${}_5$/CFT${}_4$
correspondence described by quantum mechanical theories, obtained by zooming into 
the sector near unitarity bounds of $\CN=4$ SYM on $\R \times S^3$.
Indeed, these limits can in fact be thought of as a non-relativistic limit \cite{Harmark:2014mpa}, since the relativistic magnon dispersion relation \cite{Beisert:2005tm} of the ${\cal{N}}=4$ spin chain
exhibits non-relativistic features in the SMT limit \cite{Harmark:2008gm}. 
One is thus led to consider SMT  as a  concrete and well-defined framework to formulate a holograpic correspondence involving non-relativistic string theory  and corresponding non-relativistic bulk geometries. 

The first step towards uncovering this connection was taken in Ref.~\cite{Harmark:2017rpg}, showing that strings moving in a certain type of non-relativistic target spacetime geometry, described
by a non-relativistic world-sheet action, are related to the SMT limits of the AdS/CFT correspondence. The approach taken was to first consider a target space null reduction
of the relativistic Polyakov action with fixed momentum of the string along the null isometry, leading to a covariant action for strings moving in a torsional Newton--Cartan (TNC) geometry%
 \footnote{As will be clear in Sec.~\ref{sec:TNC_Polyakov} we find in this paper that the TNC geometry is extended with a periodic target space direction.}
 \cite{Andringa:2010it,Christensen:2013lma,Christensen:2013rfa,Jensen:2014aia,Hartong:2014oma,Hartong:2014pma,Geracie:2015dea,Hartong:2015zia} after a Legendre transform that puts the momentum conservation off-shell.
  In a second step a further limit was performed, sending the string tension to zero while rescaling the Newton--Cartan clock 1-form, so as to keep the string action finite. This gives a non-relativistic world-sheet sigma-model 
describing a non-relativistic string moving in a non-relativistic geometry, which was dubbed $U(1)$-Galilean geometry. 
When applied to strings on AdS${}_5 \times S^5$ this scaling limit is realized by the SMT limits of \cite{Harmark:2008gm,Harmark:2014mpa}. The SMT limit is in turn closely related to limits of spin chains 
\cite{Kruczenski:2003gt}  (see also  \cite{Kruczenski:2004kw,Hernandez:2004uw,Stefanski:2004cw,Bellucci:2004qr,Bellucci:2006bv}) that have been studied in connection to integrability in AdS/CFT. In particular,  the simplest example of the non-relativistic world-sheet theory describes a covariant version of the Landau-Lifshitz sigma-model which is the continuum limit of the ferromagnetic $XXX_{1/2}$ Heisenberg spin chain.

In this paper we will present  a general Polyakov-type formulation
for the non-relativistic string action on TNC geometry as well as the corresponding non-relativistic sigma-model theory obtained from the SMT scaling limit. In fact, as also discussed in the present work, the formulation originally presented in \cite{Harmark:2017rpg} corresponds rather to a Nambu--Goto-type description%
\footnote{This Nambu--Goto form was also obtained in \cite{Kluson:2018egd}.}
  of these non-relativistic string theories. As part of this, we find that the correct interpretation of the $\eta$ field found in \cite{Harmark:2017rpg} is that of a periodic target space direction on which the string has a winding mode.
  
 Using the Polyakov-type formulation we show that  
the action for strings moving in TNC geometry has a local Weyl symmetry. Interestingly, after
taking the scaling limit, we find that the novel class of non-relativistic sigma models exhibits a non-relativistic version of this local Weyl symmetry. When going to a flat space gauge on the world-sheet,  we then show that the action 
possesses a symmetry corresponding to the (two-dimensional) Galilean Conformal Algebra (GCA)%
\footnote{The GCA was also observed in earlier work on non-relativistic limits of AdS/CFT \cite{Bagchi:2009my}. See also Ref.~\cite{Bagchi:2009pe} for useful work on representations of the GCA and aspects of non-relativistic conformal two-dimensional field theories.}%
, paralleling the Virasoro algebra of relativistic string theory.  Thus our novel class of non-relativistic sigma models, including those that appear in SMT limits of the AdS/CFT correspondence, represent a realization of non-relativistic conformal two-dimensional field theories. 

The existence of such a general class of non-relativistic sigma models with GCA symmetry, connected to non-relativistic strings, is expected to provide a fertile ground for further exploration. These are non-trivial interacting two-dimensional field theories that are first order in time and second order in space derivatives on the world-sheet and that couple to a type of non-relativistic target space geometry. It is interesting to note that the two-dimensional GCA also appears as the residual gauge symmetry of the tensionless  closed (super)string in the analogue of the conformal gauge  \cite{Bagchi:2013bga,Bagchi:2016yyf}.
However, the world-sheet theories in \cite{Bagchi:2013bga,Bagchi:2016yyf} appear to be of a different type.


By using the SMT limits to study the specific application to AdS/CFT of this non-relativistic sigma model, we will show that these world-sheet theories are realized in well-known physical models.  In particular, the simplest example
of the SMT/scaling limit (called the $SU(2)$ limit)  leads to a Polyakov
version of the  Landau-Lifshitz (LL) model. This shows that the LL model can
be interpreted as a non-relativistic string theory with a four-dimensional target space and, moreover, has the GCA symmetry. Part of this interpretation involves identifying 
the periodic target space direction on which the string has a winding mode
as the position on the Heisenberg spin chain. 
We also treat in detail the most general SMT limit of strings on AdS$_5 \times S^5$, called the $SU(1,2|3)$ limit. This SMT limit admits black hole solutions. Further, the other SMT limits can be viewed as special cases of the $SU(1,2|3)$ limit.

There is considerable literature on non-relativistic strings. See e.g. \cite{Gomis:2000bd,Danielsson:2000gi,Kruczenski:2003gt,Gomis:2005pg,Bagchi:2009my}
for earlier work and \cite{Andringa:2012uz,Ko:2015rha,Batlle:2016iel,Harmark:2017rpg,Morand:2017fnv,Kluson:2018egd,Bergshoeff:2018yvt,Kluson:2018grx} for more recent work).  
Interestingly, the non-relativistic string theory obtained in \cite{Gomis:2000bd,Danielsson:2000gi} from consistent low energy limits of relativistic string theory
was recently shown \cite{Bergshoeff:2018yvt} to be related to strings that couple to the so-called string NC geometry introduced in Ref.~\cite{Andringa:2012uz}.  
These formulations are intimately connected to the non-relativistic strings on TNC geometry that we discuss in this paper
and can (modulo details and subtleties) be considered to be different incarnations of the same overall structure. 
In particular,  we will show  that the Gomis-Ooguri non-relativistic string action \cite{Gomis:2000bd} can also be obtained from the action of
non-relativistic strings on TNC geometry discussed in \cite{Harmark:2017rpg} and
the present paper, by restricting the target space to flat NC spacetime. We furthermore
discuss how our formulation relates to the one in Refs.~\cite{Andringa:2012uz,Bergshoeff:2018yvt}.

For clarity, we include some words on nomenclature. The non-relativistic string theory on TNC geometry (and hence also the theory of \cite{Gomis:2000bd}) is non-relativistic in the sense that the strings move in a non-relativistic target space geometry. The actual world-sheet theory is still relativistic, and governed by a 
two-dimensional CFT.  On the other hand, the theory obtained after the  scaling limit does not only have a non-relativistic target space, but is also governed by non-relativistic
world-sheet symmetries, leading to the GCA as remarked above. 


\section{Strings on torsional Newton--Cartan geometry}
\label{sec:TNCstring}

\subsection{Polyakov action for strings on TNC geometry}
\label{sec:TNC_Polyakov}

We consider a $(d+2)$-dimensional space-time with a null isometry. One can always put the metric in the following {\sl null-reduced} form
\begin{equation}
\label{nullredmetric}
ds^2 = G_{MN} dx^M dx^N= 2 \tau (du-m) + h_{\mu\nu} dx^\mu dx^\nu\,,
\end{equation}
with $\partial_u$ being a null Killing vector field and 
where $M=(u,\mu)$, $\tau=\tau_\mu dx^\mu$ and $m=m_\mu dx^\mu$ with $x^\mu$ coordinates on a $(d+1)$-dimensional manifold. The rank-$d$ symmetric tensor $h_{\mu\nu}$ has signature $(0,1,\ldots,1)$. We assume that $u$ is a non-compact direction.
The tensors $\tau_\mu$, $m_\mu$ and $h_{\mu\nu}$ in the line element  \eqref{nullredmetric} are independent of $u$ and exhibit a set of local 
symmetries corresponding to Galilean (or Milne) boosts and
a $U(1)$ gauge transformation, along with  $(d+1)$-dimensional diffeomorphisms.
Thus these fields and their transformations correspond to 
 those of torsional Newton--Cartan (TNC) geometry \cite{Andringa:2010it,Jensen:2014aia,Hartong:2014pma,Hartong:2015zia,Geracie:2015dea} in agreement with the known fact that null reductions give rise to TNC geometry \cite{Duval:1984cj,Duval:1990hj,Julia:1994bs,Christensen:2013rfa}.

On this background, 
the Polyakov Lagrangian of a relativistic string is given by
\begin{equation}
\label{eq:P}
\mathcal{L} =-\frac{T}{2}\sqrt{-\gamma}\gamma^{\alpha\beta}\bar h_{\alpha\beta}-T\sqrt{-\gamma}\gamma^{\alpha\beta}\tau_\alpha\partial_\beta X^u\,,
\end{equation}
where $\gamma_{\alpha\beta}$ is the world-sheet metric with $\gamma$ its determinant, $X^M=X^M(\sigma^0,\sigma^1)$ the embedding coordinates of the string with $\sigma^\alpha$, $\alpha=0,1$ the world-sheet coordinates and we have performed pullbacks of the target-space fields to the world-sheet, {\sl e.g.}~$\tau_\alpha=\partial_\alpha X^\mu \tau_\mu$. We have also defined
\begin{equation}
\bar{h}_{\mu\nu} = h_{\mu\nu}-\tau_\mu m_\nu -m_\mu \tau_\nu\,, 
\end{equation}
which is invariant under local Galilean boosts. 
We are considering a closed string without winding, hence $X^M (\sigma^0,\sigma^1+2\pi)=X^M(\sigma^0,\sigma^1)$. The world-sheet has the topology of a cylinder. The world-sheet momentum current of the string's momentum in the $u$ direction is
\begin{equation}\label{eq:Pu_current}
P_u^\alpha=\frac{\partial\mathcal{L}}{\partial\partial_\alpha X^u}=-T\sqrt{-\gamma}\gamma^{\alpha\beta}\tau_\beta\,.
\end{equation}
The total momentum along $u$ is
\begin{equation}
\label{eq:Pdef}
P = \int_0^{2\pi} d\sigma^1 P_u^{0} \,.
\end{equation}
In the above formulation, the conservation of the momentum is on-shell. 

Our goal is to find an action for a closed string on the TNC geometry given by $\tau_\mu$, $m_\mu$ and $h_{\mu\nu}$. For this reason, we focus on a sector of fixed null (light-cone) momentum $P \neq 0$.%
\footnote{Note that this is analogous to the procedure of getting the action for a non-relativistic point-particle on TNC geometry via null-reduction of a massless particle \cite{Duval:1984cj,Duval:1990hj,Festuccia:2016caf}. Also in that case one works in a sector of fixed null momentum. This momentum becomes the mass of the non-relativistic particle.} It is therefore 
 convenient to find a dual formulation in which the conservation of $P$ is implemented off-shell.
To this end consider the Lagrangian%
\begin{equation}\label{eq:P2}
\mathcal{L}_{\eta A}=-\frac{T}{2}\sqrt{-\gamma}\gamma^{\alpha\beta}\bar h_{\alpha\beta}-T \Big(\sqrt{-\gamma}\gamma^{\alpha\beta}\tau_\alpha - \epsilon^{\alpha\beta} \partial_\alpha \eta \Big) A_\beta \,,
\end{equation}
where we have introduced the new fields $\eta$ and $A_\alpha$ on the world-sheet and we employ the convention $\epsilon^{01}=-\epsilon_{01}=1$ for the two-dimensional Levi-Civita symbols. This Lagrangian is classically equivalent 
to \eqref{eq:P}. This is seen by first solving the equation of motion (EOM) for $\eta$, giving that $\epsilon^{\alpha\beta} \partial_\alpha A_\beta = 0$. This is solved by $A_\alpha = \partial_\alpha \chi$ where $\chi$ is a scalar on the world-volume. We get back the Lagrangian \eqref{eq:P} by demanding no winding, i.e. identifying $X^u = \chi$.

We notice that the two components of the field $A_\alpha$ act as Lagrange multipliers in \eqref{eq:P2}. They impose the constraints
\begin{equation}
\label{eq:con}
\sqrt{-\gamma}\gamma^{\alpha\beta}\tau_\beta = -\epsilon^{\alpha\beta} \partial_\beta \eta \,.
\end{equation}
Using \eqref{eq:Pu_current}, these constraints imply $P_u^{\alpha} = T \epsilon^{\alpha\beta} \partial_\beta \eta$ and hence the conservation of the momentum current $\partial_\alpha P_u^{\alpha}=0$. Thus, we see that with the Lagrangian \eqref{eq:P2} the conservation of $P$ is implemented off-shell.

The field $\eta$ can be interpreted as the embedding field for a target-space direction $v$ dual to $u$.
We emphasize that the above rewriting of the world-sheet Lagrangian does not correspond to a T-duality since $u$ is a non-compact null-direction and since we work in a sector of fixed momentum $P$.\footnote{We will see below that some properties of the classically dual descriptions obtained by integrating out either $\eta$ or $A_\alpha$ in \eqref{eq:P2} are reminiscent of the Ro\v cek--Verlinde procedure for T-duality \cite{Rocek:1991ps}.} Note that $\eta=X^v$ needs to have non-zero winding to account for the non-zero momentum $P$ along $u$. To this end, write
\begin{equation}
\label{eq:eta_per}
\eta(\sigma^0,\sigma^1) = \frac{P}{2\pi T} \sigma^1 + \eta_{\rm per} (\sigma^0,\sigma^1)\,,
\end{equation}
where $ \eta_{\rm per}$ is periodic $ \eta_{\rm per}(\sigma^0,\sigma^1+2\pi) =  \eta_{\rm per} (\sigma^0,\sigma^1)$. As we are in a sector with fixed $P$, we interpret Eq.~\eqref{eq:eta_per}  as the target space direction $v$ being periodic with period $P/T$ and the string winding one time around $v$ (assuming here for simplicity $P>0$). 
The momentum along $v$ is zero
\begin{equation}
\label{P_tildeu}
P_v=\int_{0}^{2\pi} d\sigma^1 \frac{\partial\mathcal{L}_{\eta A}}{\partial\partial_0\eta}=T \int_{0}^{2\pi} d\sigma^1  A_1 = 0\,,
\end{equation}
using the $\eta$ EOM which tells us that $A_1 = \partial_1 X^u$ and using that $X^u$ is periodic under $\sigma^1 \rightarrow \sigma^1 + 2\pi$, which follows from the fact that the string has no winding along the $u$ direction before the change of  variables to $\eta$ and $A_\alpha$.

We now consider another form for the Lagrangian \eqref{eq:P2}. To this end, we introduce the world-sheet zweibein $e_\alpha {}^a$ and its inverse
\begin{equation}
\label{eq:inverse_zwei}
e^\alpha {}_a = \frac{1}{e} \epsilon^{\alpha\beta}  e_\beta {}^b \epsilon_{ba}\,,
\end{equation}
where $a,b=0,1$ are flat indices and $e =  \epsilon^{\alpha\beta} e_\alpha {}^0 e_\beta {}^1$. Write the world-sheet metric and its inverse as
\begin{equation}
\label{eq:zwei}
\gamma_{\alpha\beta} = \eta_{ab} e_\alpha {}^a e_\beta {}^b {} \spa \gamma^{\alpha\beta} = \eta^{ab} e^\alpha {}_a e^\beta {}_b\,.
\end{equation}
This gives $\sqrt{-\gamma} = e$. The constraints are now equivalent to
%
\begin{equation}
\epsilon^{\alpha\beta} \partial_\beta \eta = - e\, \eta^{ab} e^\alpha {}_a e^\beta {}_b \tau_\beta\,.
\end{equation}
Using $\epsilon^{\alpha\beta} = e\,  \epsilon^{ab} e^\alpha {}_a e^\beta {}_b$ this is equivalent to
\begin{equation}
\epsilon^{ab} e^\beta {}_b \partial_\beta \eta = - \eta^{ab} e^\beta {}_b \tau_\beta\,.
\end{equation}
This corresponds to the two equations $e^\alpha {}_1 \partial_\alpha \eta = e^\alpha {}_0 \tau_\alpha$ and $e^\alpha {}_0 \partial_\alpha \eta = e^\alpha {}_1 \tau_\alpha$. Using \eqref{eq:inverse_zwei} and adding and subtracting the two equations one gets the equivalent relations
\begin{equation}
\label{eq:newcon}
\epsilon^{\alpha\beta} ( e_\alpha {}^0+e_\alpha {}^1 ) ( \tau_\beta + \partial_\beta \eta ) = 0 \spa \epsilon^{\alpha\beta} ( e_\alpha {}^0-e_\alpha {}^1 ) ( \tau_\beta - \partial_\beta \eta ) = 0\,.
\end{equation}
These equations are equivalent to the constraints \eqref{eq:con}. 

Consider the field redefinition
\begin{equation} 
\label{A_redef}
A_\alpha =  m_\alpha + \frac{1}{2}(\lambda_+ - \lambda_-) e_\alpha {}^0  +\frac{1}{2} (\lambda_+ + \lambda_-) e_\alpha {}^1 \,.
\end{equation}
With this, we trade the two independent components of $A_\alpha$ for $\lambda_\pm$. Inserting this in \eqref{eq:P2} gives 
%
\begin{eqnarray}
\label{eq:P3}
\mathcal{L}_{\rm Pol} = -\frac{T}{2}&\Big[&2 \epsilon^{\alpha\beta} m_\alpha \partial_\beta \eta + e\,  \eta^{ab} e^\alpha{}_a e^\beta{}_b h_{\alpha\beta}   
 \nn \\ &&
+\lambda_+ \epsilon^{\alpha\beta} ( e_\alpha {}^0+e_\alpha {}^1 ) ( \tau_\beta + \partial_\beta \eta )
  + \lambda_- \epsilon^{\alpha\beta} ( e_\alpha {}^0-e_\alpha {}^1 ) ( \tau_\beta - \partial_\beta \eta )  \Big]  \,.
\end{eqnarray}
This is our proposal for a Polyakov-type Lagrangian for a string moving in the TNC geometry given by $\tau_\mu$, $m_\mu$ and $h_{\mu\nu}$. Apart from the TNC geometry the target space-time has one additional compact direction $v$ along which the string has a winding mode. The Lagrangian has the two Lagrange multipliers $\lambda_\pm$ that impose the constraints \eqref{eq:newcon}. Thus, also for this action the conservation of the current $P_u^\alpha$ is off-shell. 

The Lagrangian \eqref{eq:P3} is of Polyakov type and has a local Lorentz/Weyl symmetry
\begin{equation}
\label{eq:weyl}
e_\alpha {}^0+e_\alpha {}^1 \rightarrow f_+ (e_\alpha {}^0+e_\alpha {}^1) \spa e_\alpha {}^0-e_\alpha {}^1 \rightarrow f_- (e_\alpha {}^0-e_\alpha {}^1) \spa \lambda_\pm \rightarrow \lambda_\pm / f_\pm\,,
\end{equation}
for any functions $f_\pm$ on the world-sheet. For $f_-=f_+$ this constitutes a world-sheet Weyl transformation and for $f_-=f_+^{-1}$ a local Lorentz boost. Under this symmetry the world-sheet metric transforms as $\gamma_{\alpha\beta}\rightarrow f_+ f_- \gamma_{\alpha\beta}$.
It follows that the Lagrangian \eqref{eq:P3} is a two-dimensional conformal field theory.

For the Lagrangian \eqref{eq:P3} the momentum current along $v$ is
\begin{equation}
\label{eq:Ptildeu_current}
P_v^\alpha = \frac{\partial \mathcal{L}_{\rm Pol}}{\partial \partial_\alpha \eta} = T \epsilon^{\alpha\beta} A_\beta\,,
\end{equation}
with $A_\alpha$ given in \eqref{A_redef}. Using this one finds that the total momentum along $v$ is zero as in Eq.~\eqref{P_tildeu} using again the fact that $A_1 = \partial_1 X^u$ where $X^u$ is periodic.

The constraints \eqref{eq:newcon} imply that $e_\alpha {}^0 \pm e_\alpha {}^1 = h_\pm ( \tau_\alpha \pm \partial_\alpha \eta)$ where $h_\pm$ are arbitrary functions on the world-sheet.  If we substitute these expressions back into \eqref{eq:P3}, the functions $h_{\pm}$ drop out and we obtain\footnote{Note that using the local Lorentz/Weyl symmetry \eqref{eq:weyl}, we can anyway set $h_{\pm}=1$, which means choosing a gauge where $e_\alpha {}^0 = \tau_\alpha$ and $e_\alpha {}^1= \partial_\alpha \eta$.}
\begin{equation}\label{eq:NG}
\mathcal{L}_{\rm NG}=T \left(-\epsilon^{\alpha\beta}m_\alpha\partial_\beta\eta + \frac{\epsilon^{\alpha\alpha'}\epsilon^{\beta\beta'}\left(\partial_{\alpha'}\eta\partial_{\beta'}\eta-\tau_{\alpha'}\tau_{\beta'}\right)}{2\epsilon^{\gamma\gamma'}\tau_{\gamma} \partial_{\gamma'}\eta}h_{\alpha\beta}\right) \,.
\end{equation}
This Lagrangian was previously found in \cite{Harmark:2017rpg} to describe strings on a TNC geometry given by $\tau_\mu$, $m_\mu$ and $h_{\mu\nu}$. We have shown here that it is the Nambu--Goto version%
\footnote{Similar observations were made in \cite{Kluson:2018egd}.}
of the Polyakov-type Lagrangian \eqref{eq:P3}, in the sense that the two Lagrangians \eqref{eq:P3} and \eqref{eq:NG} are classically equivalent and that to get \eqref{eq:NG} from \eqref{eq:P3} we have to integrate out the world-sheet zweibein. Moreover, we have found an interpretation of $\eta$ as the embedding function of the string on a compact target-space direction $v$. Since the action is invariant under a constant shift of $\eta$, the target space direction $v$ is an isometry. Furthermore, it is wrapped by the string and thus we should regard $v$ as a compact spacelike direction. We stress that $v$ is not part of the TNC target space geometry but should rather be viewed as an additional target space dimension added to the $x^\mu$ directions of the TNC manifold.

\subsection{Relation to the Gomis--Ooguri non-relativistic string action}
\label{sec:GO}

It is possible to relate \eqref{eq:P3} to the Gomis--Ooguri non-relativistic string action \cite{Gomis:2000bd}. To do that we specialize to the flat (conformal) gauge $e_\alpha^a=\delta_\alpha^a$ and to flat TNC spacetime by choosing $m_\mu=0$, $\tau_\mu=\delta_\mu{}^0$, $h_{00}=h_{0i}=0$ and $h_{ij}=\delta_{ij}$, $i,j=1,...,d$. Setting $\eta=X^v$ we find from \eqref{eq:P3}\,,
\begin{equation}
\CL=-\frac{T}{2}\Big[ -\partial_0 X^{i}\partial_0 X^{i}+ \partial_1 X^{i}\partial_1 X^{i} 
 +\lambda_+(\partial_1-\partial_0) (X^0+X^v)+\lambda_- (\partial_1+\partial_0) (X^0-X^v)  \Big]\,.
\end{equation}
Now define
\begin{equation}
\gamma=X^0-X^v \, , \quad \bar{\gamma}=X^0+X^v\, , \quad  \beta = \pi T \lambda_- \, , \quad \bar{\beta} = \pi T \lambda_+ \,,
\end{equation}
and, after Wick rotating to the Euclidean section by $\sigma^0=-i \sigma^2$, define furthermore
\begin{equation}
z=\sigma^1+i\sigma^2\, , \quad \bar z=\sigma^1-i\sigma^2\, , \quad \partial=\frac{1}{2}\left(\partial_1-i\partial_2\right)\, , \quad \bar\partial=\frac{1}{2}\left(\partial_1+i\partial_2\right)\, .
\end{equation}
We then find the Euclidean action
\begin{equation}
\label{S_z}
S= \frac{1}{2\pi} \int d^2 z\left( 2\pi T \, \partial X^{i }\bar\partial X^{i}+\beta \bar \partial \gamma+ \bar{\beta} \partial \bar{\gamma} \right)\,,
\end{equation}
which agrees with the Gomis--Ooguri action (Eq. (3.8) of \cite{Gomis:2000bd}, omitting the instantonic part). 

We are thus able to match the relativistic string with a fixed momentum along a non-compact null direction with the Gomis--Ooguri string action. This was possible by performing a duality transformation whereby the non-compact null direction $u$ is replaced by a compact spacelike direction $v$. The importance of a compact target space direction in the context of non-relativistic string theory was also emphasized in \cite{Danielsson:2000gi}. The Gomis--Ooguri string action was obtained in \cite{Gomis:2000bd} by performing a large speed of light limit of the relativistic string in the background of a Kalb--Ramond 2-form. Here we thus see that there is an entirely different way in which the same result can be obtained. 

It is shown in Appendix \ref{app:NCstring} that the Lagrangian \eqref{eq:P3} can be mapped to the non-relativistic Polyakov action for a string moving in what is called a string Newton--Cartan geometry \cite{Bergshoeff:2018yvt} under certain conditions. In Appendix \ref{app:NCstring} it will be also shown that contact with the work of \cite{Bergshoeff:2018yvt} requires that we impose $d\tau=0$. String Newton--Cartan geometry is essentially Newton--Cartan geometry with one additional direction - here denoted by $v$. In \cite{upcoming} this conection will be explored further by including the Kalb--Ramond 2-form and the dilaton.




\section{Strings with  non-relativistic world-sheet theories \label{sec:NRstring}
}

The string theories discussed in the previous section are non-relativistic from the target space-time perspective. However, the world-sheet theories are still described by relativistic CFTs. In this section we will go one step further and take a scaling limit whereby also the world-sheet theory becomes a non-relativistic field theory. We will show that one can obtain first order time derivative sigma models describing strings moving in a target space-time that is closely related to TNC geometry. These sigma models will be shown to admit infinite dimensional symmetry algebras. These are the analogue of the classical infinite dimensional algebra of local conformal symmetries of the relativistic string theory.

\subsection{Polyakov action for strings with non-relativistic world-sheet theories} 
\label{sec:NR_Polyakov}

In \cite{Harmark:2017rpg} a zero tension limit of the Lagrangian \eqref{eq:NG} (combined with a limit on the target space geometry to ensure finiteness of the action in the limit) was introduced in which one obtains a string with a non-relativistic world-sheet theory. In the following we shall generalize this scaling limit to our Polyakov-type Lagrangian \eqref{eq:P3}.

Our starting point is the Lagrangian \eqref{eq:P3} for a string on a TNC geometry described by $\tau_\mu$, $m_\mu$ and $h_{\mu\nu}$ with one extra dimension added - the compact isometry parametrized by $v$. 
In terms of the pullbacks, the scaling limit of \cite{Harmark:2017rpg} is 
\begin{equation}
\label{eq:scaling1}
c \rightarrow \infty \spa T = \frac{\tilde{T}}{c} \spa \tau_\alpha = c^2 \tilde{\tau}_\alpha  \spa m_\alpha = \tilde{m}_\alpha \spa h_{\alpha\beta} = \tilde{h}_{\alpha\beta} \spa \eta = c\, \tilde{\eta}\, , 
\end{equation}
where the tilde quantities are the rescaled ones. 
Note that $P$ in \eqref{eq:Pdef} does not scale.
We supplement this with the following scaling of the zweibeins and Lagrange multipliers
\begin{equation}
\label{eq:scaling2}
e_\alpha {}^0 = c^2 \tilde{e}_\alpha {}^0 \spa e_\alpha {}^1 = c\, \tilde{e}_\alpha {}^1 \spa \lambda_\pm = \frac{\omega}{2c^3} \pm \frac{\psi}{2c^2}\,.
\end{equation}
Note that $e = c^3 \tilde{e}$, $e^\alpha {}_0 = c^{-2} \tilde{e} ^\alpha {}_0$ and $e^\alpha {}_1 = c^{-1} \tilde{e} ^\alpha {}_1$ under this scaling. Clearly, this scaling limit is a non-relativistic limit on the world-sheet as it scales $e_\alpha {}^0$ and $e_\alpha {}^1$ differently. It is essentially a limit in which we are sending the world-sheet speed of light to infinity.

Taking the scaling limit \eqref{eq:scaling1}-\eqref{eq:scaling2} of \eqref{eq:P3} one gets the Lagrangian
\begin{equation}
\label{eq:NRP}
\mathcal{L}_{\rm NRPol} = - \frac{T}{2} \left[   2 \epsilon^{\alpha\beta} m_\alpha \partial_\beta \eta + e \, e^{\alpha} {}_1 e^{\beta}{}_1 h_{\alpha\beta}  + \omega \epsilon^{\alpha\beta} e_\alpha {}^0 \tau_\beta  + \psi \epsilon^{\alpha\beta} \Big( e_\alpha{}^0 \partial_\beta \eta + e_{\alpha}{}^1 \tau_\beta \Big) \right]\,,
\end{equation}
where to avoid heavy notation we have removed the tildes from the tension and all the tilded fields in \eqref{eq:scaling1} and \eqref{eq:scaling2}. As we shall see below, this is the Lagrangian for a string with a non-relativistic world-sheet theory propagating in a non-relativistic target space whose geometric properties differ from those of TNC geometry in a manner to be  discussed below. 
Note that the scaling limit of the zweibeins \eqref{eq:scaling2} is consistent with the definition of the inverse zweibeins \eqref{eq:inverse_zwei}. Thus this formula can be used also after the scaling limit.

The rescaling of the $\eta$ field is consistent with the scaling of the tension in \eqref{eq:scaling1} such that also after the scaling we have
\begin{equation}
\label{eq:eta_per2}
\eta(\sigma^0,\sigma^1) = \frac{P}{2\pi T} \sigma^1 + \eta_{\rm per} (\sigma^0,\sigma^1)\,,
\end{equation}
where $ \eta_{\rm per}$ is periodic $ \eta_{\rm per}(\sigma^0,\sigma^1+2\pi) =  \eta_{\rm per} (\sigma^0,\sigma^1)$. Again, we can interpret $\eta$ as the embedding map for a periodic target space direction $\tilde v$ with period $P/T$ so that $\eta= X^{\tilde v}$. To avoid clutter we now also drop the tilde on $v$.
Eq.~\eqref{eq:eta_per2} means then that the closed string is winding one time around the periodic $v$ direction. 
The momentum current along $v$ is
\begin{equation}
\label{eq:NR_v_current}
P_{v}^\alpha = \frac{\partial \mathcal{L}_{\rm NRPol}}{\partial \partial_\alpha \eta}= T \epsilon^{\alpha\beta} A_\beta \spa A_\alpha =  m_\alpha + \frac{1}{2} \psi \, e_\alpha {}^0 \,.
\end{equation}
This can also be obtained by taking the scaling limit \eqref{eq:scaling1}-\eqref{eq:scaling2} of \eqref{eq:Ptildeu_current} and \eqref{A_redef}.
From the EOM of $\eta$ this is seen to be conserved $\partial_\alpha P_{v}^\alpha=0$ which gives that $A_\alpha$ is closed. Indeed from  Section \ref{sec:TNCstring} we have that $A_\alpha = \partial_\alpha X^u$ where $X^u$ is periodic under $\sigma^1 \rightarrow \sigma^1 + 2\pi$ and this is unaffected by the scaling limit \eqref{eq:scaling1}-\eqref{eq:scaling2}. From this we get that the momentum along $v$ is again zero
\begin{equation}
\label{eq:NR_Pv}
P_{v} = \int_0 ^{2\pi} d\sigma^1 P_{v}^0 = T \int_0 ^{2\pi} d\sigma^1 \partial_1 X^u = 0\,.
\end{equation}

The action \eqref{eq:NRP} is invariant under local transformations that act on 
the target space fields $\tau_\mu$, $m_\mu$ and $h_{\mu\nu}=\delta_{ab}e^a_\mu e^b_\nu$ as \cite{Harmark:2017rpg} 
\begin{equation}\label{eq:limitTNCsym}
\delta\tau_\mu=0\,,\quad\delta m_\mu = \partial_\mu\sigma\,,\quad\delta h_{\mu\nu} =2 \tau_{(\mu} e^a_{\nu)}\lambda_a\,. 
\end{equation}
along with target space diffeomorphisms. The former symmetries were shown in 
\cite{Harmark:2017rpg} to correspond to the gauging of a spacetime symmetry algebra consisting of a direct sum of the (massless) Galilei algebra $\texttt{Gal}$ and a $U(1)_\sigma$. This is analogous to the 
 way that gauging the Bargmann algebra (massive Galilei algebra) gives TNC geometry \cite{Andringa:2012uz,Hartong:2015zia}. 
The resulting geometry was dubbed $U(1)$-Galilean geometry. We note that the $U(1)$ factor in this algebra is crucial in order to allow for time derivatives in the theory. A massless Galilei symmetry without such a $U(1)$ only admits spatial derivatives.\footnote{One way to see this is to note that for a free particle, massless Galilean symmetries, imply that the dispersion relation is $\vec k^2=0$ where $\vec k$ is the momentum vector. This follows from the Bargmann dispersion relation $\omega=\frac{\vec k^2}{2m}$ after setting $m=0$ or from the relativistic massless dispersion relation $\omega=c\vert\vec k\vert$ after sending $c\rightarrow\infty$.} 


The scaling limit \eqref{eq:scaling1}-\eqref{eq:scaling2} respects the reduction of \eqref{eq:P3} to the Nambu--Goto type Lagrangian \eqref{eq:NG}. Indeed, if one puts $e_\alpha {}^0 = \tau_\alpha$ and $e_\alpha {}^1= \partial_\alpha \eta$  (where the fields are not rescaled) we get \eqref{eq:NG} and we can subsequently take the scaling limit \eqref{eq:scaling1} to obtain
\begin{equation}
\label{eq:NRNG}
\mathcal{L}_{\rm NRNG} = -T \left[   \epsilon^{\alpha\beta} m_\alpha \partial_\beta \eta 
+  \frac{\epsilon^{\alpha\alpha'}\epsilon^{\beta\beta'}\tau_{\alpha'}\tau_{\beta'}}{2\epsilon^{\gamma\gamma'}\tau_{\gamma} \partial_{\gamma'}\eta}h_{\alpha\beta} \right]\,.
\end{equation}
This is the Lagrangian found in  \cite{Harmark:2017rpg}. Alternatively, one can first take the scaling limit of \eqref{eq:P3} to obtain \eqref{eq:NRP}, and then subsequently solve for the Lagrange multipliers $\omega$ and $\psi$ as we shall see below. 

The Lagrangian \eqref{eq:NRP} is invariant under the local symmetry
\begin{equation}
\label{eq:NRweyl}
e_\alpha {}^0 \rightarrow f   e_\alpha {}^0 \spa e_\alpha {}^1\rightarrow f  e_\alpha {}^1  + \hat{f}  e_\alpha {}^0 \spa \omega \rightarrow \frac{1}{f} \omega - \frac{\hat{f}}{f^2} \psi \spa \psi \rightarrow \frac{1}{f}\psi\,,
\end{equation}
for arbitrary functions $f$ and $\hat{f}$ on the world-sheet. These constitute local Galilei/Weyl symmetries acting on the world-sheet vielbeins and Lagrange multipliers. The Lagrange multipliers $\omega$ and $\psi$ in the Lagrangian \eqref{eq:NRP} impose the constraints
\begin{equation}
\epsilon^{\alpha\beta} e_\alpha {}^0 \tau_\beta=0 \spa  \epsilon^{\alpha\beta} \Big( e_\alpha{}^0 \partial_\beta \eta + e_{\alpha}{}^1 \tau_\beta \Big) = 0\,.
\end{equation}
The general solution to these constraints is
\begin{equation}
e_\alpha {}^0 = h \, \tau_\alpha \spa e_\alpha {}^1 = h \, \partial_\alpha \eta + \hat{h} \, \tau_\alpha\,,
\end{equation}
where $h$ and $\hat{h}$ can be any functions on the world-sheet. In particular, we can use the local Galilei/Weyl symmetries \eqref{eq:NRweyl} to set $e_\alpha {}^0 = \tau_\alpha$ and $e_\alpha {}^1= \partial_\alpha \eta$. One obtains again \eqref{eq:NRNG}, so this shows in particular that the Polyakov-type Lagrangian \eqref{eq:NRP} is equivalent to the Nambu--Goto type Lagrangian \eqref{eq:NRNG} originally found in \cite{Harmark:2017rpg}, in analogy to what was done before the large $c$ limit.

The world-sheet geometry in the Polyakov-type formulation is described by $e_\alpha {}^0$ and $e_\alpha {}^1$. The Lagrange multipliers $\omega$ and $\psi$ can be written as 
\begin{equation}
\omega=e^\alpha{}_0 \chi_\alpha\,,\qquad\psi=e^\alpha{}_1 \chi_\alpha\,,
\end{equation}
where $e^\alpha{}_0$ and $e^\alpha{}_1$ are the inverse vielbeins. The latter transform as
\begin{equation}
 e^\alpha{}_0\rightarrow f^{-1}  e^\alpha{}_0  - \frac{\hat{f}}{f^2}  e^\alpha{}_0 \spa e^\alpha {}_1 \rightarrow f^{-1}   e^\alpha{}_1\,.
\end{equation}
Hence in order to reproduce the transformations of $\omega$ and $\psi$ in \eqref{eq:NRweyl} we do not need to transform $\chi_\alpha$. Thus, even though $\chi_\alpha$ is a world-sheet 1-form, we do not think of it as being part of the geometry (like we would for example in the case of TNC geometry where $m$ transforms under the local Galilean boosts that act on the vielbeins). In other words the world-sheet geometry can be thought of as a 2-dimensional massless Galilei geometry, i.e. the geometry obtained by gauging the massless 2-dimensional Galilei algebra.

\subsection{GCA symmetry of the non-relativistic sigma-model}

Combining the local Galilei/Weyl symmetry \eqref{eq:NRweyl} with local diffeomorphisms we have enough symmetry to transform the zweibeins to the gauge\footnote{We can use one diffeomorphism and the Weyl transformation to set $e_\alpha{}^0=\delta_\alpha^0$. We can then use the local Galilean boost to set $e_\alpha{}^1\propto \delta_\alpha^1$ and subsequently the second diffeomorphism to set $e_\alpha{}^1=\delta_\alpha^1$.}
\begin{equation}
\label{eq:flat_gauge}
e_\alpha {}^a = \delta_\alpha ^a \,.
\end{equation}
In this flat gauge, the Lagrangian \eqref{eq:NRP} takes the form
\begin{equation}
\label{eq:NRPflat}
\mathcal{L}_{\rm flat} = - \frac{T}{2} \left[  2 m_\mu \epsilon^{\alpha\beta} \partial_\alpha X^\mu  \partial_\beta \eta +h_{\mu\nu} \partial_1 X^\mu \partial_1 X^\nu    + \omega \tau_\mu \partial_1 X^\mu    + \psi  \Big(  \partial_1 \eta - \tau_\mu \partial_0 X^\mu  \Big) \right]\,.
\end{equation}
We remind the reader that the $\mu$ index does not include the $v$ direction.
The residual gauge transformations, i.e. those obtained from the local Galilei/Weyl transformations and world-sheet diffeomorphisms preserving the flat gauge \eqref{eq:flat_gauge} are
\begin{equation}\label{eq:residual1}
\sigma^0 \rightarrow \hat{\sigma}^0 (\sigma)= F(\sigma^0) \spa \sigma^1 \rightarrow \hat{\sigma}^1 (\sigma)= F'(\sigma^0) \sigma^1 + G(\sigma^0 ) \,,
\end{equation}
for any functions $F(\sigma^0)$ and $G(\sigma^0)$ with $F(\sigma^0)$ monotonically increasing. While $X^\mu(\sigma)$ and $\eta(\sigma)$ transform as scalars, $\omega(\sigma)$ and $\psi (\sigma)$ transform as
\begin{equation}\label{eq:residual2}
\omega(\sigma)= F' \hat{\omega} (\hat{\sigma}) + (F'' \sigma^1 + G') \hat{\psi} (\hat{\sigma})
\spa
\psi (\sigma)= F' \hat{\psi} (\hat{\sigma})\,.
\end{equation}
It is straightforward to verify that \eqref{eq:residual1} and \eqref{eq:residual2} leave the flat gauge Lagrangian \eqref{eq:NRPflat} invariant.

The transformations \eqref{eq:residual1} are generated by the following infinitesimal diffeomorphisms
\begin{equation}\label{eq:diffeo}
\xi^0=f(\sigma^0)\,,\qquad\xi^1=g(\sigma^0)+f'(\sigma^0)\sigma^1\,.
\end{equation}
 Assuming $f$ and $g$ to be analytic so that they can be analytically continued to the complex plane we can perform a Laurent expansion of $f$ and $g$.
 Let us define
\begin{equation}
f=-\sum_n a_n(\sigma^0)^{n+1}\,,\qquad g=\sum_n b_n (\sigma^0)^{n+1}\,,
\end{equation} 
and if we furthermore define
\begin{equation}
\xi^\alpha\partial_\alpha=\sum_n\left(a_n L_n+b_nM_n\right)\,,
\end{equation}
then from \eqref{eq:diffeo} we find the algebra generators $L_n$ and $M_n$
\begin{eqnarray}
L_n & = & -(\sigma^0)^{n+1}\partial_0-(n+1)(\sigma^0)^n\sigma^1\partial_1\,,\\
M_n & = & (\sigma^0)^{n+1}\partial_1\,.
\end{eqnarray}
They satisfy the algebra
\begin{equation}
\left[L_n\,,L_m\right]=(n-m)L_{n+m}\,,\qquad\left[L_n\,,M_m\right]=(n-m)M_{n+m}\,.
\end{equation}
This algebra is the 2-dimensional Galilei conformal algebra (GCA) without any central extensions. The latter are absent because we treated the world-sheet theory classically. We thus conclude that the non-relativistic sigma model \eqref{eq:NRPflat} has the GCA as its infinite dimensional symmetry algebra.

It is known that the GCA can be obtained  from two copies of the Virasoro algebra
via a contraction \cite{Bagchi:2009my}. Here, we have found a general class of non-relativistic sigma models exhibiting this symmetry. It would be very interesting to study in more detail whether the GCA symmetry of these sigma-models plays the same role as the Virasoro algebra (including its central extension) in relativistic string theory.


\section{Spin Matrix theory limits of strings on AdS$_5\times S^5$
\label{sec:SMT}}

In \cite{Harmark:2017rpg} it was found that the Spin Matrix theory limits introduced in \cite{Harmark:2014mpa} are realizations of the zero tension scaling limit \eqref{eq:scaling1} of the Nambu--Goto-type Lagrangian \eqref{eq:NG} that gives the Lagrangian \eqref{eq:NRNG}. In this section we generalize this statement to the scaling limit \eqref{eq:scaling1}-\eqref{eq:scaling2} of the Polyakov-type Lagrangian \eqref{eq:P3} that gives the Lagrangian \eqref{eq:NRP} with a non-relativistic Weyl symmetry. 

In the course of this, we find a particularly nice interpretation of the limit of the compact $v$ direction as the position of the spins in the spin-chain limit of Spin Matrix theory.

\subsection{Spin Matrix theory}

Let us first briefly review the Spin Matrix theory (SMT) limits \cite{Harmark:2014mpa} of the AdS/CFT correspondence.%
\footnote{See also \cite{Harmark:2008gm} as well as \cite{Harmark:2006di,Harmark:2006ta,Harmark:2007px}. On SMT-type limit related to $\ads_3$/CFT$_2$ correspondence see \cite{Hartong:2017bwq}.}
The AdS/CFT correspondence asserts a duality between $SU(N)$ $\CN=4$ super Yang-Mills theory (SYM) and type IIB string theory on $\ads_5\times S^5$. On the gauge theory side,  $\CN=4$ SYM on $\R \times S^3$ with gauge group $SU(N)$ has certain unitarity bounds that we schematically can write as
\begin{equation}
\label{unibound}
E \geq Q \,,
\end{equation}
where for a given state $E$ is the energy and $Q$ is a linear sum over the Cartan charges of $PSU(2,2|4)$ being the two angular momenta $S_1$ and $S_2$ on $S^3$ and the three R-charges $J_1$, $J_2$ and $J_3$. We have set the radius of $S^3$ to one.
Taking a limit \cite{Harmark:2014mpa}
\begin{equation}
\label{SMT_limit}
\lambda \rightarrow 0\spa N = \mbox{fixed} \spa \frac{E-Q}{\lambda} = \mbox{fixed} \,,
\end{equation}
$\CN=4$ SYM simplifies greatly and is effectively described by a quantum mechanical theory called Spin Matrix theory (SMT). SMTs are quantum mechanical theories characterized by having a Hilbert space made from harmonic oscillators with both an index in a representation of a Lie group (called the spin group) as well as matrix indices corresponding to the adjoint representation of $SU(N)$ (or $U(N)$) subject to a singlet constraint for the matrix indices. The Hamiltonian of SMT is factorized into a spin and a matrix part, acting on the Hilbert space by removing two excitations and creating two new ones \cite{Harmark:2014mpa}.
One can equivalently take the limit in the grand canonical ensemble by approaching a zero-temperature critical point. 

In the limit \eqref{SMT_limit} only states in $\CN=4$ SYM on $\R \times S^3$ for which $E$ is close to $Q$ will survive. The rest of the states decouple. Writing $E = E_0 + \lambda E_1 + \CO(\lambda^2)$ where $E_0$ is the classical (tree-level) energy and $E_1$ is the one-loop correction, we see that only states with classical energy $E_0=Q$ survive. This gives the Hilbert space of the resulting SMT. Moreover, the Hamiltonian of the resulting SMT corresponds to the one-loop correction $E_1$ in $\CN=4$ SYM. In fact, we write
\begin{equation}
\label{SMT_limit2}
H = Q + g \lim_{\lambda\rightarrow 0} \frac{E-Q}{\lambda} = Q + g E_1 \,,
\end{equation}
as the Hamiltonian of SMT where $g$ is the coupling of the SMT. The global symmetry group $PSU(2,2|4)$ of $\CN=4$ SYM reduces to the so-called spin group in the SMT limit. In table \ref{tab:SMT} we have listed five limits of $\CN=4$ SYM where $E=Q$ defines a supersymmetric subsector of $\CN=4$ SYM, which means that SMT describes a near-BPS regime of $\CN=4$ SYM. 

\begin{table}[ht]
\begin{center}
\begin{tabular}{|c||c|c|c|c|}
\hline Unitarity bound $E\geq Q$   & $G_s$  & Cartan diagram  & $R_s$   & $d+2$ \\  
\hline 
$Q = J_1+J_2$ & $SU(2)$ & $\bigcirc$ & $[1]$  & 4 \\
\hline
$Q= J_1+J_2+J_3$ & $SU(2|3)$ & $\bigcirc \!\!-\!\!\!-\!\! \textstyle \bigotimes  \!\!-\!\!\!-\!\! \bigcirc \!\!-\!\!\!-\! \bigcirc$ &$[0,0,0,1]$ & 6 \\
\hline
$Q=S_1+J_1+J_2$ & $SU(1,1|2)$ & $\textstyle \bigotimes \!\!-\!\!\!-\!\! \bigcirc \!\!-\!\!\!-\!\! \textstyle \bigotimes$ & $[0,1,0]$ &  6\\
\hline
$Q=S_1+S_2+J_1$ & $SU(1,2|2)$ & $\bigcirc \!\!-\!\!\!-\!\!  \textstyle \bigotimes \!\!-\!\!\!-\!\!
\bigcirc \!\!-\!\!\!-\!\! \textstyle \bigotimes
$ & $[0,0,0,1]$ &  6 \\
\hline
$Q=S_1+S_2+J_1+J_2+J_3$ & $SU(1,2|3)$ & $\bigcirc \!\!-\!\!\!-\!\!  \textstyle \bigotimes \!\!-\!\!\!-\!\!
\bigcirc \!\!-\!\!\!-\!\! \bigcirc \!\!-\!\!\!-\!\! \textstyle \bigotimes$
 &$[0,0,0,1,0]$ &  10 \\
\hline
\end{tabular}
\caption{Five unitarity bounds of $\CN=4$ SYM that give rise to SMTs in the limit \eqref{SMT_limit}. For each limit we list the spin group $G_s$, the Cartan diagram for the corresponding algebra and the representation $R_s$ of the algebra (in terms of Dynkin labels) that defines the Spin Matrix Theory for a given limit. Moreover, $d+2$ is the space-time dimension of the target space for the corresponding sigma-model (see Section \ref{sec:SMT_AdS_limit}).
 \label{tab:SMT}} 
\end{center}
 \end{table}
 
In the planar limit $N \rightarrow \infty$ SMT reduces to a nearest-neighbor spin chain \cite{Harmark:2014mpa}. For the five cases of table \ref{tab:SMT} the spins of the spin chain are in the representation $R_s$ for the algebra of the spin group $G_s$. Since the spin chain Hamiltonian defines uniquely the SMT also for finite $N$ one can think of SMT as a finite-$N$ generalization of nearest-neighbor spin chains.

The low energy excitations of spin chains are magnons. The dispersion relation of a single magnon in $\CN=4$ SYM is \cite{Beisert:2005tm}
\begin{equation}
\label{eq:disp1}
E-Q = \sqrt{1+\frac{\lambda}{\pi^2} \sin^2 \frac{p}{2}}-1 \,,
\end{equation}
where $p$ is the momentum. Taking the SMT limit \eqref{SMT_limit2} gives \cite{Harmark:2008gm}
\begin{equation}
\label{eq:disp2}
H-Q = \frac{g}{2\pi^2} \sin^2 \frac{p}{2} \,.
\end{equation}
This corresponds to a dispersion relation of a magnon in a nearest-neighbor spin chain.
The magnon excitations dominate for $g \gg 1$ which one can think of as the strong coupling limit of SMT.
Comparing \eqref{eq:disp1} and \eqref{eq:disp2} we see that the SMT limit clearly can be thought of as a non-relativistic limit \cite{Harmark:2014mpa}. This is even clearer for the small momentum limit $p\ll 1$, corresponding to the pp-wave limit, in which one goes from a relativistic dispersion relation to a Galilean one \cite{Harmark:2008gm}. Thus, one observes that the SMT limits in this sense correspond to non-relativistic limits of $\CN=4$ SYM. 

Using coherent states on each spin chain site, one can find a semi-classical limit with many magnon states on the spin chain. In the limit in which the spin chain is long, one can furthermore find a long wavelength limit. This procedure results in a sigma-model description in a semi-classical limit of the spin chain \cite{Kruczenski:2003gt} (see also \cite{Kruczenski:2004kw,Hernandez:2004uw,Stefanski:2004cw,Bellucci:2004qr,Bellucci:2006bv} and \cite{Harmark:2008gm}). The sigma-model is based on the coset related to the representation of the spin group in Table \ref{tab:SMT}.
The resulting sigma-models are what we below in Section \ref{sec:SMT_AdS_limit} will find from taking SMT limits of strings on $\ads_5\times S^5$.

Consider the $SU(2)$ SMT in the planar limit $N =\infty$. This is described by the XXX$_{1/2}$ ferromagnetic Heisenberg spin-chain. In this case $Q=J_1+J_2$ is the length of the spin chain. For $Q \gg 1$ and in a semi-classical regime, one obtains an effective sigma-model description of the spin chain called the Landau-Lifshitz model. The Landau-Lifshitz model is
\begin{equation}
\label{eq:LLmodel}
\CL_{\rm LL} = \frac{Q}{4\pi} \left[ \cos \theta \dot{\phi} - \frac{1}{4} \Big( (\theta')^2 + \sin^2 \theta (\phi')^2 \Big)\right] \,,
\end{equation}
where the fields $\theta$ and $\phi$ are functions of $\sigma^0$ and $\sigma^1$. The fields are periodic under $\sigma^1 \rightarrow \sigma^1 + 2\pi$. We also defined $\dot{\phi} = \partial_0 \phi$ and $\phi'=\partial_1 \phi$.
In the semi-classical limit $Q\rightarrow \infty$ that gives the Lagrangian \eqref{eq:LLmodel} one identifies $\sigma^1$ with the position on the spin chain \cite{Kruczenski:2003gt} (also reviewed in~\cite{Harmark:2008gm})
\begin{equation}
\label{eq:LLmodel_site}
\sigma^1 = 2\pi \frac{k}{Q} \,,
\end{equation}
where $k$ is a site on the spin chain and $Q$ is the length of the spin chain, thus explaining that the fields are periodic in $\sigma^1$ with period $2\pi$. Since the spin chain description arises from single-trace operators \cite{Minahan:2002ve}, the cyclicity of the trace gives that the total momentum along the spin chain is zero, corresponding to the condition
\begin{equation}
\label{eq:LLmodel_Pzero}
\int_0^{2\pi}  d\sigma^1 \cos\theta \phi' = 0 \,,
\end{equation}
in the semi-classical limit.


\subsection{Limits of strings on $\ads_5\times S^5$}
\label{sec:SMT_AdS_limit}

We now turn to the string theory side of the AdS/CFT correspondence. Using the dictionary of the AdS/CFT correspondence, we can formulate a SMT limit \eqref{SMT_limit} of $\CN=4$ SYM as a limit of type IIB string theory on $\ads_5\times S^5$. As we shall see, this corresponds to a limit of the sigma-model of the string that realizes the $c\rightarrow \infty$ scaling limit \eqref{eq:scaling1}-\eqref{eq:scaling2}.

We write the metric on $\ads_5\times S^5$ in the global patch as
\begin{equation}
\label{eq:metads_R}
ds^2 = R^2 \Big[ - \cosh^2 \rho \, dt^2 + d\rho^2 + \sinh^2 \rho \, d\Omega_3^2 + d\Omega_5^2 \Big] \,.
\end{equation}
The AdS/CFT dictionary states that $4\pi g_s = \lambda /N$ and $R/l_s = \lambda^{1/4}$ where 
$\lambda$ is the 't Hooft coupling of the gauge theory and $g_s$ and $l_s$ are the string coupling and string length of the string theory side, respectively, and $R$ is radius of $S^5$ and $\ads_5$. On the string theory side $N$ translates to the flux of the self-dual Ramond-Ramond five-form field strength on $\ads_5$ and on $S^5$.
Moreover, we can translate the unitarity bounds \eqref{unibound} with $Q$ given in Table \ref{tab:SMT} into corresponding BPS bounds $E \geq Q$ where $E$ is the energy corresponding to the global time coordinate $t$, $S_1$ and $S_2$ are the angular momenta on the $S^3$ within $\ads_5$ and $J_1$, $J_2$, $J_3$ are the angular momenta on $S^5$, all measured in units of $1/R$. 
%
%

In terms of the metric \eqref{eq:metads_R} the string tension is $1/(2\pi l_s^2)$. However, since the factor $R^2$ in \eqref{eq:metads_R} is uniform we can include it in the tension instead of the metric.
With this,  one gets an effective string tension
\begin{equation}
T = \frac{R^2}{2\pi l_s^2} = \frac{ \sqrt{4\pi g_s N} }{2\pi} \,.
\end{equation}
In using this as the string tension, we should rescale the metric \eqref{eq:metads_R} as
\begin{equation}
\label{eq:metads}
ds^2 = - \cosh^2 \rho \, dt^2 + d\rho^2 + \sinh^2 \rho \, d\Omega_3^2 + d\Omega_5^2  \,.
\end{equation}

For a given BPS bound $E \geq Q$ (with $Q$ of Table \ref{tab:SMT}) the SMT limit of type IIB string theory on $\ads_5 \times S^5$ is
\begin{equation}
\label{string_SMT_limit}
g_s \rightarrow 0 \spa N = \mbox{fixed} \spa \frac{E-Q}{g_s} = \mbox{fixed} \,.
\end{equation}
This is the translation of the limit \eqref{SMT_limit} to the string theory side. 
To implement this, we use that given a particular BPS bound defined by $Q$ of Table \ref{tab:SMT} one can find coordinates $u$, $x^\mu$, $y^I$ for the metric \eqref{eq:metads} for $\ads_5\times S^5$, where $\mu= 0,1,...,d$, $I=1,2,...,2n$ and $d=8-2n$, such that
\begin{itemize}
\item $\partial_{x^0}$ and $\partial_u$ are  Killing vector fields with 
\begin{equation}
\label{eq:Pformula}
i \partial_{x^0}  = E - Q  \spa P = - i \partial_u = \frac{1}{2} ( E + Q ) \,,
\end{equation}
where $P$ is defined as the momentum along $u$.
\item For $y^I =0$ one has $g_{uu}=0$ and one can furthermore put the metric restricted to $y^I =0$ in a form
\eqref{nullredmetric} with
\begin{equation}
m_0=h_{00} = h_{0i} = 0 \,,
\end{equation}
for $i=1,2,...,d$.
\end{itemize}
Here $n$ is the number of angular momenta (out of $S_1$, $S_2$, $J_1$, $J_2$ and $J_3$) that are not included in the unitarity bound. This gives $2n$ directions $y^I$, that we call {\sl external directions}, that realize $n$ rotation planes associated to the $n$ commuting angular momenta that are not included. In the SMT limit \eqref{string_SMT_limit} one has a confining potential with effective mass proportional to $1/g_s$ for each of these $n$ rotation planes that drives the strings to sit at the minimum of the potential located at $y^I = 0$. This gives an effective reduction of the number of spatial dimensions in the SMT limit for four out of the five limits listed in Table \ref{tab:SMT}. In Table \ref{tab:SMT} we have furthermore recorded the number of surviving space-time dimensions $d+2= 10 -2n$ for each case. 

To make contact with the scaling limit \eqref{eq:scaling1}-\eqref{eq:scaling2}, we identify
\begin{equation}
\label{eq:SMT_c}
c = \frac{1}{\sqrt{4\pi g_s N}} \,,
\end{equation}
and scale the $x^0$ coordinate as
\begin{equation}
\label{eq:SMT_x0}
x^0 = c^2 \tilde{x}^0 \,,
\end{equation}
while the coordinates $u$, $x^i$, $i=1,...,d$, are held fixed in the $c\rightarrow \infty$ limit. Writing
\begin{equation}
\tau = \tau_\mu dx^\mu = F dx^0 + \beta_i dx^i= c^2 F d\tilde{x}^0 + \beta_i dx^i \,,
\end{equation}
before the limit, one sees that in the $c\rightarrow \infty$ limit the first term goes like $c^2$ which means $\tau_\alpha$ has the correct scaling. Taking the limit one finds
\begin{equation}
\tau =  F d\tilde{x}^0 \,,
\end{equation}
where we removed the tildes on the LHS. 

In addition to $\tau$ having the correct scaling, one finds also that $m_\alpha$ and $h_{\alpha\beta}$ do not scale in the $c\rightarrow \infty$ limit. As seen in section \ref{sec:TNCstring}, $\eta$ scales like $\eta=c\tilde{\eta}$ and $u$ is held fixed. Thus, in this way the SMT limit \eqref{string_SMT_limit} is a realization of the scaling limit \eqref{eq:scaling1}. This should be supplemented with the zweibein scalings \eqref{eq:scaling2} when we do not fix a gauge for the zweibeins before the limit. 
Below we carry out this limit more explicitly in two of the cases of table \ref{tab:SMT}. These are the same cases considered in \cite{Harmark:2017rpg}.

In the $c\rightarrow \infty$ limit described above the effective string tension becomes 
\begin{equation}
\label{eq:Tlimit}
T= \frac{1}{2\pi} \,.
\end{equation}
Using \eqref{eq:Pformula} we see that the momentum $P= - i\partial_u$ along the $u$ direction becomes
\begin{equation}
\label{eq:Plimit}
P = Q \,.
\end{equation}
%



\subsection{$SU(2)$ limit and Polyakov Lagrangian for Landau-Lifshitz model}

Our first example is the SMT/scaling limit  towards the BPS bound $E\geq Q=J_1+J_2$. This has $n=3$ and $d=2$.
Our starting point is the metric \eqref{eq:metads} for $\ads_5\times S^5$. Write the five-sphere part as
\begin{equation}
d\Omega_5^2 = d \alpha^2 + \sin^2 \alpha \, d\beta^2 + \cos^2 \alpha \Big[ (d\Sigma_1)^2 + (d\gamma + A)^2 \Big]   \,,
\end{equation}
with
\begin{equation}
(d\Sigma_1)^2 = \frac{1}{4} ( d\theta^2 + \sin^2 \theta d\phi^2 ) \spa A = \frac{1}{2} \cos\theta d\phi \,.
\end{equation}
We have $E = i \partial_t$ and $Q = -i \partial_\gamma$. Write now $t$ and $\gamma$ as linear functions of $x^0$ and $u$. To satisfy~\eqref{eq:Pformula} we need%
\footnote{Note here that since $\gamma$ is compact  the $(x^0,u)$ plane is in fact a strip. This does not affect the non-compactness of $\tilde{x}^0$ (defined below) after the limit.}
\begin{equation}
t = x^0 - \frac{1}{2} u \spa \gamma = x^0 + \frac{1}{2} u \,.
\end{equation}
This, in turn, ensures that $g_{uu}=0$ for $\rho=\alpha=0$.
Using this, one can write the metric \eqref{eq:metads} of AdS$_5\times S^5$ as 
\begin{eqnarray}
\label{eq:adsmet1}
 ds^2 &=&\cos^2 \alpha  \Big[  2 \tau ( du - m ) + h_{\mu\nu} dx^\mu dx^\nu \Big]
  - (\sinh^2 \rho + \sin^2 \alpha) \left(dx^0-\frac{1}{2} du\right)^2 \nn \\ &&+ 
 d\rho^2+ \sinh^2\rho \, d\Omega_3^2 + d\alpha^2 + \sin^2 \alpha \, d\beta^2 \,,
\end{eqnarray}
with $d+2=4$ and
\begin{equation}
\label{eq:LLmh}
\tau= dx^0 + \frac{1}{4}\cos \theta d\phi \spa m = - \frac{\cos \theta}{2}  d\phi  \spa h_{\mu\nu}dx^\mu dx^\nu = \frac{1}{4} ( d\theta^2 + \sin^2 \theta d\phi^2 )  \,.
\end{equation}
These coordinates for AdS$_5\times S^5$ can be seen to correspond to the above-mentioned $(u,x^\mu,y^I)$ coordinates by identifying $x^\mu=(x^0,\theta,\phi)$ and finding a coordinate transformation for the external directions between the six coordinates given by $\rho$, $\alpha$, $\beta$ and the coordinates for the three-sphere to $y^I$, $I=1,2,...,6$, such that $y^I=0$ corresponds to $\rho=\alpha=0$. The six external directions have a potential proportional to $(\sinh^2\rho + \sin^2\alpha)/g_s$ that confines them to the point $\rho=\alpha=0$ corresponding to $y^I=0$ in the SMT limit \eqref{string_SMT_limit} \cite{Harmark:2008gm}. Therefore we set $\rho=\alpha=0$ in the following.

The SMT/scaling limit \eqref{eq:scaling1}-\eqref{eq:scaling2} combined with \eqref{string_SMT_limit}, \eqref{eq:SMT_c} and \eqref{eq:SMT_x0} then gives the sigma-model Lagrangian \eqref{eq:NRP} with the $U(1)$-Galilean background given by
\begin{equation}
\label{eq:target_su2}
\tau= d\tilde{x}^0 \spa m = - \frac{\cos \theta}{2}  d\phi  \spa h_{\mu\nu}dx^\mu dx^\nu = \frac{1}{4} ( d\theta^2 + \sin^2 \theta d\phi^2 )  \,.
\end{equation}
In addition to the three directions $x^\mu=(\tilde{x}^0,\theta,\phi)$, the four-dimensional target space includes the compact direction $v$ as well. 
The closed string on this background has winding number one along $v$ as in \eqref{eq:eta_per2} with $\eta=X^v$. Using \eqref{eq:eta_per2}, \eqref{eq:Tlimit} and \eqref{eq:Plimit} we find
\begin{equation}
\label{eq:eta_per_su2}
\eta(\sigma^0,\sigma^1) = Q \sigma^1 + \eta_{\rm per} (\sigma^0,\sigma^1) \,,
\end{equation}
as we shall see below this has an interesting interpretation. Note that the $v$ direction has period $2\pi Q$.

Fixing the gauge on the world-sheet to $e_\alpha {}^0=\tau_\alpha$ and $e_\alpha {}^1 = \partial_\alpha \eta$, and furthermore choosing the following gauge for $\eta$
\begin{equation}
\label{eq:eta_gauge}
\eta (\sigma^0, \sigma^1 ) = Q \sigma^1 \,,
\end{equation}
as well as the static gauge choice
\begin{equation}
X^0 (\sigma^0,\sigma^1)= Q^2 \sigma^0 \,,
\end{equation}
one finds using \eqref{eq:NRNG} that the sigma-model reduces to the Landau-Lifshitz model \eqref{eq:LLmodel}. 

We consider now the condition \eqref{eq:NR_Pv} of zero momentum along $v$.
The momentum current along $v$ is given by \eqref{eq:NR_v_current}. Since $e_1{}^0 = \tau_\mu \partial_1 X^\mu = 0$ we find $P^0_v = T m_\mu \partial_1 X^\mu$. Therefore, the condition of zero momentum along $v$ is
\begin{equation}
\int_0^{2\pi} d\sigma^1 \cos \theta \partial_1 \phi  =0 \,.
\end{equation}
This is seen to correspond to the zero momentum condition \eqref{eq:LLmodel_Pzero} for the Landau-Lifshitz sigma-model.

The above shows that the general Lagrangian \eqref{eq:NRP} on the background \eqref{eq:target_su2} and \eqref{eq:eta_per_su2} is an equivalent description of the Landau-Lifshitz model. This general description can be interpreted as a Polyakov version of the Landau-Lifshitz model in which we can interpret the Landau-Lifshitz model as a non-relativistic string theory with a non-relativistic four-dimensional target space. 

As part of this, we see that the compact $v$ direction is identified with the position on the Heisenberg spin chain. This is seen by combining \eqref{eq:LLmodel_site}, \eqref{eq:eta_per_su2} and \eqref{eq:eta_gauge}. This shows that in the general situation in which we do not gauge-fix $\eta=X^v$, the compact $v$ direction gives the position on the spin chain. In detail, one has that a given site $k$ on the spin chain is identified with $v/(2\pi)$. 

The identification of $v$ with the position on the spin chain is further verified by the connection to the requirement of zero momentum along the $v$ direction in the general description of Section \ref{sec:NR_Polyakov}. The origin of this is the fact that the $u$ direction is a null isometry and hence cannot be periodic and have winding, as explained in Section \ref{sec:TNC_Polyakov}. Obviously, this is in particular the case when starting with the $\ads_5\times S^5$ background.  The condition of having zero winding along $u$ leads to zero momentum along the $v$ direction, both before and after the $c\rightarrow \infty$ limit \eqref{eq:scaling1}-\eqref{eq:scaling2}. This is seen to perfectly correspond to the fact that the spin chain description also dictates a zero momentum condition along the periodic spin chain, in accordance with the identification of $v$ with the position on the spin chain.

\subsection{$SU(1,2|3)$ limit of strings on $\ads_5\times S^5$}

The most interesting SMT limit of the AdS/CFT correspondence is the $SU(1,2|3)$ case of Table \ref{tab:SMT} with
\begin{equation}
\label{eq:Q123}
Q = S+ J \spa S=S_1+S_2\spa J=J_1+J_2+J_3 \,.
\end{equation}
Taking the SMT limit \eqref{SMT_limit} of $\CN=4$ SYM one gets $SU(1,2|3)$ SMT. This is the SMT with the largest possible Hilbert space and global symmetry group (of the ones obtained as limits of $\CN=4$ SYM). Moreover, the other four SMTs of Table \ref{tab:SMT} can be obtained as subsectors of the $SU(1,2|3)$ SMT, both with respect to the Hilbert space and the Hamiltonian. Connected to this is the fact that the $SU(1,2|3)$ SMT limit is the only SMT limit with $n=0$, hence one does not decouple any directions when taking the SMT limit on the string theory side. This means that the resulting target-space geometry is ten-dimensional. Finally, it is also interesting to note that the $1/16$ BPS supersymmetric black hole in $\ads_5\times S^5$ of \cite{Gutowski:2004yv} obeys the BPS bound $E=Q$ and it survives thus the $SU(1,2|3)$ SMT limit on the string theory side of the correspondence. 


Below we consider the SMT/scaling limit towards the BPS bound $E \geq Q$ with $Q$ given by \eqref{eq:Q123}. The starting point is the metric \eqref{eq:metads} for $\ads_5\times S^5$. We parametrize the three- and five-sphere as
\begin{equation}
d\Omega_5^2 = (d\Sigma_2 )^2 + ( d\psi + A)^2 \spa d\Omega_3^2 = (d\Sigma_1)^2 + (d\chi + C)^2 \,,
\end{equation}
where $(d\Sigma_k)^2$ are Fubini-Study metrics for $\C P^k$, $k=1,2$, given by
\begin{equation}
(d\Sigma_2)^2 = d\xi_1^2 + \frac{1}{4} \sin^2 \xi_1 ( d\xi_2^2 + \sin^2 \xi_2 d\zeta_1^2 ) + \frac{1}{4} \sin^2 \xi_1 \cos^2 \xi_1 ( d\zeta_2 + \cos \xi_2d\zeta_1 )^2 \,,
\end{equation}
\begin{equation}
(d\Sigma_1)^2 = \frac{1}{4} ( d\xi_3^2 + \sin^2\xi_3 d\zeta_3^2 ) \,.
\end{equation}
with the one-forms
\begin{equation}
A = - \frac{1}{2} \cos \xi_2 \sin^2 \xi_1 d\zeta_1 + \left( \frac{1}{3} - \frac{1}{2} \sin^2 \xi_1 \right) d\zeta_2
\spa C = \frac{1}{2} \cos \xi_3 d\zeta_3  \,.
\end{equation}
We have
\begin{equation}
E = i \partial_t \spa S = - i \partial_\chi \spa J = - i \partial_\psi \,.
\end{equation}
We write now $t$, $\chi$ and $\psi$ as linear functions of $x^0$, $u$ and a new variable $w$ that is periodic with period $2\pi$. Then Eq.~\eqref{eq:Pformula} fixes the coefficients in front of $x^0$ and $u$.%
\footnote{One can derive the coefficients of $x^0$ and $u$  by demanding only $i\partial_{x^0} = E - Q$, $g_{uu}=0$ and that $w$ is periodic. From $i\partial_{x^0} = E - Q$ we get $t=x^0 + b_1 u + c_1 w$, $\chi=x^0 + b_2 u + c_2 w$ and $\psi = x^0 + b_3 u + c_3 w$. $g_{uu}=0$ gives $b_1^2=b_2^2=b_3^2$. Demanding periodicity of $\chi$, $\psi$ and $w$ gives that $b_1 = -b_2=-b_3$. Assuming $P> 0$ we find $-b_1=b_2=b_3>0$. We choose $b_3=\frac{1}{2}$.}
 Demanding furthermore that $-i \partial_w = S$ we get
\begin{equation}
t= x^0 - \frac{1}{2}u \spa \chi=x^0 + \frac{1}{2}u +w \spa \psi = x^0+ \frac{1}{2}u \,.
\end{equation}
This brings the metric of $\ads_5\times S^5$ in the null-reduced form \eqref{nullredmetric} with
\begin{equation}
\tau = \cosh^2\rho\, dx^0 + \frac{1}{2}  \sinh^2 \rho ( dw + C )  + \frac{1}{2}A  \spa m =  - \tanh^2 \rho ( dw + C ) - \cosh^{-2}\rho \, A \,,
\end{equation}
\begin{equation}
h_{\mu\nu}dx^\mu dx^\nu = d\rho^2+  \tanh^2 \rho (dw + C-A)^2  + \sinh^2\rho \, (d\Sigma_1)^2 + (d\Sigma_2)^2 \,,
\end{equation}
corresponding to an $(8+1)$-dimensional TNC background with isometry group $SU(1,2|3)$.

We take the SMT/scaling limit \eqref{eq:scaling1}-\eqref{eq:scaling2} combined with \eqref{string_SMT_limit}, \eqref{eq:SMT_c} and \eqref{eq:SMT_x0}. This gives the sigma-model Lagrangian \eqref{eq:NRP}  with the $(8+1)$-dimensional $U(1)$-Galilean background given by
\begin{equation}
\label{eq:su123_back1}
\tau = \cosh^2\rho\, d\tilde{x}^0   \spa m =  - \tanh^2 \rho ( dw + C ) - \cosh^{-2}\rho \, A \,,
\end{equation}
\begin{equation}
\label{eq:su123_back2}
h_{\mu\nu}dx^\mu dx^\nu = d\rho^2+  \tanh^2 \rho (dw + C-A)^2  + \sinh^2\rho \, (d\Sigma_1)^2 + (d\Sigma_2)^2 \,,
\end{equation}
and with $2\pi T=1$.
One can check that this background has global isometry group $SU(1,2|3)$ as well. The full target space geometry is ten-dimensional, as the $(8+1)$-dimensional $U(1)$-Galilean geometry is supplemented by the compact $v$ direction with period $2\pi Q$. The closed string has winding number one on this background, so that $\eta(\sigma^0,\sigma^1) = Q \sigma^1 + \eta_{\rm per} (\sigma^0,\sigma^1)$ where $\eta_{\rm per}$ is periodic in $\sigma^1$.

Choosing the gauge $e_\alpha {}^0=\tau_\alpha$ and $e_\alpha {}^1=\partial_\alpha \eta$ for the world-sheet zweibeins and the gauge $\eta(\sigma^0,\sigma^1)=Q \sigma^1$ and $X^0(\sigma^0,\sigma^1)=Q^2 \sigma^0$ for the target space embedding, we obtain the Lagrangian
\begin{equation}
\CL = - \frac{Q}{2\pi} \left( m_\mu \partial_0 X^\mu + \frac{1}{2} h_{\mu\nu} \partial_1 X^\mu \partial_1 X^\nu \right) \,,
\end{equation}
with $m_\mu$ and $h_{\mu\nu}$ given in \eqref{eq:su123_back1}-\eqref{eq:su123_back2}. The zero-momentum condition is
\begin{equation}
\int_0^{2\pi} d\sigma^1 m_\mu \partial_1 X^\mu = 0 \,.
\end{equation}
To have a semi-classical sigma-model one needs $Q$ to be large.

\section{Discussion}

One of the main results of this paper is that we have presented a Polyakov-type formulation (see eq.~\eqref{eq:NRP}) of the non-relativistic world-sheet sigma-model action, which was recently \cite{Harmark:2017rpg} found in Nambu--Goto form. This action is obtained from a  scaling limit of a string action describing strings with Poincar\'e world-sheet symmetry but moving in a non-relativistic (TNC) target spacetime, for which we  also have 
obtained the corresponding Polyakov-type formulation (see eq.~\eqref{eq:P3}). Another central result of the paper is that this new non-relativistic world-sheet sigma model has the following properties: 
\begin{itemize}
\item The target space is a type of non-relativistic geometry, namely $U(1)$-Galilean geometry  \cite{Harmark:2017rpg}, extended with a periodic target space direction. 
\item The residual gauge symmetry in flat gauge is the Galilean Conformal Algebra. 
\end{itemize} 
Thus this novel class of non-relativistic sigma-models, describing non-relativistic conformal two-dimensional field theories, provides an interesting setting for further exploration of non-relativistic string theory.

Importantly, the non-relativistic world-sheet sigma-model action is concretely realized in the SMT limits \cite{Harmark:2014mpa}  of the AdS/CFT correspondence. The latter are tractable limits of the  AdS/CFT correspondence,
and we thus obtain a covariant form for the corresponding non-relativistic string theories, being well-defined and quantum mechanically consistent. Moreover, following the two bullets above, our description i) elucidates the nature of the target space that these non-relativistic strings move in, and ii) shows that these world-sheet theories are non-relativistic two-dimensional conformal field theories. 
The SMT sigma-models of Section~\ref{sec:SMT} can thus be considered as specific 
string theory realizations relevant to the non-relativistic sector of quantum gravity/holography advocated in the introduction. They represent a natural starting point for further studies of the corresponding quantum theory. 

In this connection already the simplest case, being the $SU(2)$ SMT limit,
provides a new perspective on the Landau--Lifshitz sigma model, which is known to appear as the long wavelength limit of integrable spin chains \cite{Kruczenski:2003gt}. 
For this  we have shown that
the spin chain direction has an interpretation as the periodic target space direction of the non-relativistic geometry. The same holds for
more general SMT limits and the corresponding integrable theories which are generalizations of the Landau--Lifshitz sigma model \cite{Kruczenski:2004kw,Hernandez:2004uw,Stefanski:2004cw,Bellucci:2004qr,Bellucci:2006bv}. In particular, we discussed the most symmetric and most general $SU(1,2|3)$ SMT limit. 

With these sigma-models in hand, we can start to address the question
of how non-relativistic gravity (with the gauge symmetries of $U(1)$-Galilean geometry) emerges from SMT, and subsequently more generally
for the entire class of non-relativistic sigma models of Section~\ref{sec:NRstring}.
Since these models have the GCA as a symmetry group, and hence an underlying (non-relativistic) conformal symmetry, it would be very interesting
to see whether it is possible to compute the analogue of the standard beta-functions
of relativistic string theory. This would point the way towards uncovering the
underlying low-energy non-relativistic gravity theory.%
\footnote{It would furthermore be interesting to find a string theory connection to the action for
the non-relativistic limit of Einstein gravity, recently found in 
\cite{Hansen:2018ofj}.}
The backgrounds described for the Polyakov-type actions in Section~\ref{sec:SMT} are then naturally expected to be solutions of such gravity theories. 

To complete this program, one should obviously include all the
string theory background fields in the analysis performed in this paper.
The first steps towards this will be considered in an upcoming work 
\cite{upcoming}. Moreover, the inclusion of supersymmetry, D-branes\footnote{See \cite{Harmark:2016cjq} for the $SU(2)$ SMT limit of the non-abelian Born-Infeld action for D-branes in the AdS/CFT correspondence.}
and
many of the other standard features of relativistic string theory are natural extensions to consider. More generally, understanding the general properties of sigma-models with GCA symmetry would be a further important direction. 
We also note that there are tantalizing connections with doubled field theory
and doubled geometry  \cite{Ko:2015rha,Morand:2017fnv} that are worthwhile to examine. Finally, one could speculate that the non-relativistic corner of string theory and its relation to SMT could be useful towards understanding closed string field theory.

\section*{Acknowledgements}

We thank Eric Bergshoeff and Gerben Oling for useful discussions.
The work of JH is supported by the Royal Society University Research Fellowship ``Non-Lorentzian Geometry in Holography'' (grant number UF160197).
The work of TH and NO is supported in part by the project ``Towards a deeper understanding of  black holes with non-relativistic holography'' of the Independent Research Fund Denmark (grant number DFF-6108-00340).  
The research of ZY was supported in part by Perimeter Institute for Theoretical Physics. Research at Perimeter Institute is supported by the Government of Canada through the Department of Innovation, Science and Economic Development and by the Province of Ontario through the Ministry of Research, Innovation and Science.
LM thanks Niels Bohr Institute, Perugia University and INFN of Perugia for support.
JH and NO thank the Perimeter Institute for hospitality during the early phases of this work. ZY thanks Niels Bohr Institute for hospitality.

\appendix

%
%

\section{Relation to string Newton--Cartan geometry}
\label{app:NCstring}

In \cite{Bergshoeff:2018yvt} the theory for strings moving in a string-NC geometry \cite{Andringa:2012uz} was considered in detail. Being a theory of strings moving in a non-relativistic target geometry, hence akin in this respect to the one presented in Section \ref{sec:TNCstring} and Ref. \cite{Harmark:2017rpg}, it is interesting to study the precise relation between the two. We work out the precise dictionary in Section \ref{sec:BGY_rel}, specializing  to backgrounds with zero Kalb-Ramond field and dilaton\footnote{A more general analysis will be included in \cite{upcoming}.} and discuss the difference between the two frameworks. Moreover,  in \ref{sec:BGY_scal} we also consider the large $c$ scaling limit (in the spirit of section \ref{sec:NRstring}) of the action given in \cite{Bergshoeff:2018yvt} and briefly comment on the result. 

\subsection{Dictionary from String NC geometry}
\label{sec:BGY_rel}
The Polyakov Lagrangian for the sigma model describing a closed string in a string-NC $(d+2)$-dimensional geometry is \cite{Bergshoeff:2018yvt}
\begin{equation}
\label{eq:BGY}
\CL= - \frac{T}{2} \left[ \sqrt{-\gamma} \gamma^{\alpha\beta} H_{\alpha\beta} + \lambda \epsilon^{\alpha\beta} ( e_\alpha{}^0+e_\alpha{}^1) ( \tau_\beta{}^0+\tau_\beta{}^1) + \bar\lambda\epsilon^{\alpha\beta}  (e_\alpha{}^0-e_\alpha{}^1)  (\tau_\beta{}^0 -\tau_\beta{}^1)  \right]\, ,
\end{equation}
where the zweibein $e_\alpha{}^a$ is introduced as in \eqref{eq:zwei}, $\tau_\alpha{}^A$, $m_\alpha{}^A$\footnote{It is important to notice that the superscripts $0,1$ in \eqref{eq:BGY} are of different nature: those on the zweibein specify the flat world-sheet tangent directions $a$, while those on the clock form select the longitudinal directions $A$ on the target spacetime tangent space.} and $H_{\alpha\beta}$ are the pullbacks on the world-sheet of the spacetime tensors $\tau_M{}^A$, $m_M{}^A$ and
\begin{equation}
\label{eq:HMN}
H_{MN} = E_M {}^{A'} E_N {}^{B'} \delta_{A'B'} +( \tau_M {}^A m_N {}^B + \tau_N {}^A m_M{}^B) \eta_{AB}\, .
\end{equation}
The indices $A=0,1$ and $A'=2,\ldots,d+1$ respectively denote \emph{longitudinal} and \emph{transverse} directions of the manifold's tangent space. The two-dimensional foliation is specified by two one-forms $\tau_M{}^A$ which are taken to satisfy
\begin{equation}
\label{eq:tau_constr}
D_{[M}\tau_{N]}{}^A=0 \, ,
\end{equation}
where the derivative is covariant with respect to longitudinal $SO(1,1)$ Lorentz transformations, so it includes a spin-connection field $\omega_M{}^{AB}$. 

The target space symmetries of \eqref{eq:BGY} are those of string-NC geometry, given by
\begin{eqnarray}
\delta \tau_M{}^A & = & \CL_\xi \tau_M{}^A+\Lambda^A{}_B \tau_M{}^B\,,\label{eq:tautrafo}\\
\delta E_M{}^{A'} & = & \CL_\xi E_M{}^{A'}+\lambda^{A'}{}_A\tau_M{}^A+\lambda^{A'}{}_{B'}E_M{}^{B'}\,,\\
\delta m_M{}^A & = & \CL_\xi m_M{}^A-\lambda_{A'}{}^AE_M{}^{A'}+\Lambda^A{}_B m_M{}^B+D_M\sigma^A+\sigma^A{}_B\tau_M{}^B\,,
\end{eqnarray}
where the target space-time diffeomorphisms are generated by the Lie derivatives along $\xi^M$, and where $\Lambda^A{}_B=\Lambda \epsilon^A{}_B$ and $\lambda^{A'}{}_{B'}$ describes longitudinal $SO(1,1)$ and transverse $SO(d)$ transformations, respectively. The parameters $\lambda^{A'}{}_A$ describe string Galilei boost transformations and $\sigma^A$ form string Bargmann-type (non-central) extensions of the algebra. Finally, the parameteres $\sigma^A{}_B$ are only constrained to be traceless, in order that the Lagrangian \eqref{eq:BGY} remains invariant, but they will play no role in what follows. The indices $A$ and $A'$ are raised/lowered with $\delta_{AB}$ and $\delta_{A'B'}$. On the world-sheet we have diffeomorphisms, Weyl transformations and local Lorentz transformations acting on the world-sheet tangent space.

Let us split $M=(v, \mu)$ with $X^{v}=\eta$ a (spatial) longitudinal direction. We choose to set
\begin{equation}
\label{eq:tau_set}
\tau_\mu {}^0 = \tau_\mu \spa \tau_\mu {}^1 = 0 \spa \tau_{v} {}^0 = 0\spa \tau_{v} {}^1 = 1\, .
\end{equation}
The last two conditions can be shown to be a gauge choice. The relevant infinitesimal symmetry transformation here is \eqref{eq:tautrafo} with $\Lambda^A{}_B=\Lambda \epsilon^A{}_B$. We can set $\tau_{v} {}^0 = 0$ using $\Lambda$ and subsequently perform a diffeomorphism choosing $\xi^M$ such that $\partial_v \xi^\mu=0$ (ensuring preservation of the first condition). This still leaves us the freedom to set $\tau_{v}{}^1=1$. The residual gauge transformations are given by the subset of gauge transformations respecting $\delta \tau_v{}^0=\delta\tau_v{}^1=0$.

Instead, setting $\tau_\mu{}^1=0$ is not a gauge choice but rather a truncation of the theory. Indeed the residual gauge transformation of $\tau_\mu{}^1$ turns out to be
\begin{equation}
\delta \tau_{\mu}{}^1=\xi^\rho \partial_\rho \tau_{\mu}{}^1+\tau_{\rho}{}^1 \partial_\mu \xi^\rho+\xi^{v} \partial_{v} \tau_{\mu}{}^1+\partial_{\mu} \xi^{v}-\tau_{\mu}{}^0\tau_{\rho}{}^0 \partial_{v} \xi^\rho\,,
\end{equation}
and we do not have enough freedom to set $\tau_\mu{}^1=0$. Nevertheless, as will be shown in \cite{upcoming}, the situation is improved when we include the Kalb--Ramond 2-form as in that case we do not need to truncate $\tau_\mu{}^1=0$ to obtain a relation with the string NC description. In other words in the presence of the Kalb--Ramond 2-form we do not need to arbitrarily impose any conditions on $\tau_\mu{}^A$ other than the gauge choices already discussed. The residual transformations preserving \eqref{eq:tau_set} are those respecting 
\begin{equation}
\label{eq:tau_res2}
\Lambda+\tau_\mu \partial_{v} \xi^\mu=0\, , \quad \partial_{v} \xi^{v}=0\, , \quad \Lambda \tau_\mu +\partial_\mu \xi^{v}=0\, .
\end{equation}

With $X^v$ being a longitudinal direction, it makes sense to also impose $E_v{}^{A'}=0$. This is easily done through a string Galilei boost, infinitesimally given by $\delta E_M{}^{A'}=\lambda^{A'}{}_A \tau_M{}^A$. Residual transformations must then respect $E_{\mu}{}^{A'} \partial_v \xi^\mu + \lambda^{A'}{}_1=0$.

We furthermore choose 
\begin{equation}
\label{eq:mu_set}
m_\mu{}^0 =  m_\mu \, , \quad m_\mu{}^1 =  m_{v}{}^0 \tau_\mu \, , \quad m_{v} {}^1 = 0\, .
\end{equation}
The last of the above can be imposed without loss of generality (and respecting previous gauge choices) by using for example the extension transformation $\delta m_v{}^1=D_v \sigma^1$. For the combination $m_\mu{}^1-m_v{}^0 \tau_\mu$ one instead finds that similarly to $\tau_\mu{}^1$ it cannot be put to zero by gauge fixing.

From the above it follows that $H_{vv}=H_{v\mu}=0$ and $H_{\mu\nu}= h_{\mu\nu}-m_\mu \tau_\nu-m_\nu\tau_\mu$ where we defined $h_{\mu\nu} = E_\mu {}^{A'} E_\nu {}^{B'} \delta_{A'B'}$. With the redefinition
\begin{equation}
\label{eq:multip_shift}
\lambda=\lambda_+ + \frac{1}{e} \epsilon^{\alpha\beta}(e_\alpha{}^0-e_\alpha{}^1)m_\beta\, , \quad \bar\lambda=\lambda_-+\frac{1}{e} \epsilon^{\alpha\beta}(e_\alpha{}^0+e_\alpha{}^1)m_\beta\, ,
\end{equation}
one then easily verifies that \eqref{eq:BGY} becomes \eqref{eq:P3}. 

We end this section with some comments. It turns out that in order to correctly retrieve the target space NC symmetries starting from those of \eqref{eq:BGY} we also need to assume that $V=\partial_v$ is a Killing vector, i.e. the fields $\tau_\mu$, $m_\mu$ and $h_{\mu\nu}$ do not depend on $v$. Using this, an important consequence\footnote{This can be seen by putting $M=\mu$ and $N=v$ in \eqref{eq:tau_constr}.} of \eqref{eq:tau_constr} is then that $\omega_M{}^{AB}=0$. From \eqref{eq:tau_set} and \eqref{eq:tau_constr} we then obtain
\begin{equation}
\partial_{[\mu} \tau_{\nu]}=0\, .
\end{equation}
Thus we have the additional condition that $\tau_\mu$ be closed. The necessity of setting $d\tau=0$ can be seen by studying the Lagrangian \eqref{eq:BGY} in which we substitute \eqref{eq:tau_set}. Doing so one can see that whenever $m_\mu{}^1\neq 0$ the TNC transformation $\delta m_\mu =\partial_\mu \sigma$ is a symmetry if and only if $d\tau=0$. This same calculation also shows that when we set $m_\mu{}^1=0$ by hand we recover the usual TNC gauge symmetry $\delta m_\mu =\partial_\mu \sigma$ for any $\tau$.


Therefore, though we showed that the novel Polyakov string action we propose in this paper is closely related to the one in \cite{Bergshoeff:2018yvt}, a fundamental distinction on the allowed backgrounds is to be made. The relevance of this is underlined by the fact that a nonzero $d \tau$ is essential in order to link the non-relativistic string theory described in this paper to Spin Matrix Theory limits of the AdS/CFT correspondence analyzed in Section \ref{sec:SMT}. In fact one can straightforwardly verify that $d\tau \neq 0$ (before the SMT limit is taken) for all examples in section \ref{sec:SMT}.

\subsection{Scaling limit of string NC geometry sigma model}
\label{sec:BGY_scal}

We record here how to take a scaling limit of \eqref{eq:BGY}, similarly to what is done in Section \ref{sec:NRstring}. The limit is non-relativistic both on the target spacetime and on the world-sheet. For simplicity, we first redefine the Lagrange multipliers according to a more general version\footnote{This shift of the multipliers has the nice property of making them invariant under extension transformations parametrized by $\sigma^A$, i.e. $\delta_{\sigma^A} \lambda_{\pm}=0$.} of \eqref{eq:multip_shift}
\begin{equation}
\label{eq:multiplier2}
\lambda=\lambda_+ + \frac{1}{e} \epsilon^{\alpha\beta}(e_\alpha{}^0-e_\alpha{}^1)(m_\beta{}^0-m_\beta{}^1)\, , \quad \bar\lambda=\lambda_-+\frac{1}{e} \epsilon^{\alpha\beta}(e_\alpha{}^0+e_\alpha{}^1)(m_\beta{}^0+m_\beta{}^1)\, .
\end{equation}
The limit is 
\begin{equation}
\label{eq:BGY_scaling}
\begin{array}{c} \displaystyle
 c\rightarrow \infty \spa T = \frac{1}{c} \tilde{T}   \spa e_\alpha {}^0 = c^2  \, \tilde{e}_\alpha{}^0   \spa e_\alpha {}^1 = c \, \tilde{e}_\alpha{}^1 \spa \lambda_\pm = \frac{\omega}{2c^3} \pm \frac{\psi}{2c^2}\, ,\\[4mm] \displaystyle 
\tau_\alpha {}^0 = c^2 \, \tilde{\tau}_\alpha{}^0\spa \tau_\alpha {}^1 = c \, \tilde{\tau}_\alpha{}^1  \spa H^\perp_{\alpha\beta} = \tilde{H}^\perp_{\alpha\beta} \spa m_\alpha {}^0 = \tilde{m}_\alpha{}^0\spa m_\alpha {}^1 = \frac{1}{c} \tilde{m}_\alpha{}^1\, .
\end{array}
\end{equation}
The outcome is the Lagrangian 
\begin{equation}
\label{eq:BGY_largec}
\CL=-\frac{T}{2}\left[ e \, e^{\alpha}{}_1 e^{\beta}{}_1 H^\perp_{\alpha\beta} +2 \epsilon^{\alpha\beta} (\tau_\alpha{}^0  m_\beta{}^1- \tau_\alpha{}^1 m_\beta{}^0)+\omega \epsilon^{\alpha\beta} e_\alpha{}^0\tau_\beta{}^0+\psi \epsilon^{\alpha\beta} (e_\alpha{}^0 \tau_\beta{}^1+ e_\alpha{}^1 \tau_\beta{}^0)\right]
\end{equation}
where we removed tildes to avoid clutter. The inverse zweibein is still defined by \eqref{eq:inverse_zwei}. The above Lagrangian is invariant under the world-sheet local Galilei/Weyl symmetry \eqref{eq:NRweyl}, so the theory also has non-relativistic (Galilean) conformal symmetry. For the study of the full symmetry algebra that emerges from \eqref{eq:BGY_largec}, resulting from the contraction \eqref{eq:BGY_scaling} on the string Galilei algebra of Ref. \cite{Andringa:2012uz}, we refer to \cite{upcoming}.

\addcontentsline{toc}{section}{References}


\providecommand{\href}[2]{#2}\begingroup\raggedright\endgroup

\end{document}